\documentclass{JAC2000}

\usepackage{epsfig}
\input psfig.tex  
\newcommand{\beq}{\begin{equation}}
\newcommand{\eeq}{\end{equation}}
\newcommand{\bea}{\begin{eqnarray}}
\newcommand{\eea}{\end{eqnarray}}
\newcommand{\ds}{\displaystyle}
\newcommand{\bfg}{\begin{figure}}

\begin{document}
\title{ANALYSIS AND SIMULATION OF \\
THE ENHANCEMENT OF THE CSR EFFECTS}
\author{R. Li, Jefferson Lab, 12000 Jefferson Ave., Newport News, VA 23606, USA}

\maketitle

\begin{abstract}  

Recent measurements of the coherent synchrotron radiation (CSR) effects indicated that
the observed beam emittance growth and energy modulation are often bigger than previous predictions
based on Gaussian longitudinal charge distributions.
In this paper, by performing a model \,study, we show both analytically
and numerically
that when the longitudinal bunch charge distribution involves concentration
of charges in a \, small fraction of the bunch length, enhancement of the CSR
self-interaction beyond the Gaussian prediction may occur.
The level of this enhancement is sensitive to the level of the local charge 
concentration.

\end{abstract}
\vspace{-0.1in}
\section{Introduction} 

When a short bunch with high charge is transported
through a magnetic bending system, orbit-curvature-induced bunch
self-interaction  via CSR and space charge can potentially induce 
energy modulation in the bunch and cause emittance growth. 
Even though the earlier analytical results
for CSR self-interaction \cite{derbenev,murphy} based on the rigid-line-charge model
can be applied for general longitudinal charge distributions, since
the analytical results for a Gaussian beam are explicitly given,
one usually applies the  Gaussian results to predict the CSR effects
using the measured or simulated rms bunch length. Similarly, 
a self-consistent simulation \cite{li} was developed ealier to study the CSR effect
on bunch dynamics for general bunch distributions; however, the simulation
is usually carried out using an assumed initial Gaussian longitudinal phase space distribution.
Recent CSR  experiments \cite{braun,piot} indicated that the measured energy spread and emittance growth  
are sometimes bigger than previous Gaussian predictions,
especially when a bunch is fully compressed or over-compressed.
In this paper, we explore the possible enhancement of the CSR self-interaction force
due to extra longitudinal concentration of charges as opposed to a Gaussian distribution.
This study reveals a general feature of the CSR self-interaction: whenever there is
longitudinal charge concentration in a small fraction of a bunch length, enhancement
of the CSR effect beyond the Gaussian prediction can occur; moreover, the
level of this enhancement is sensitive to the level of the local charge concentration within
a bunch. This sensitivity should be given serious considertation in designs of future machines.

\section{Bunch Compression Optics}

When an electron bunch is fully compressed by a magnetic chicane,  the final bunch length 
and the inner structure of the final longitudinal phase space are determined by
many details of the machine design. In this paper, we investigate only the RF curvature effect,
which serves as a model to illustrate the possible sensitivity of the CSR interaction to the 
longitudinal charge distribution.

In order to study the CSR self-interaction for a compressed bunch, let us first
find the longitudinal charge distribution for our model bunch when it is fully compressed
by a chicane.
Consider an electron bunch with $N$ total electrons. The longitudinal charge density
of the bunch at time $t$ is 
$\rho(s,t)=Ne n(s,t)$   ($\int n(s,t)ds=1$),
 where $s$ is the distance from the reference electron, 
and $n(s,t)$ is the longitudinal density distribution of the bunch.
At $t=t_{0}$, let the bunch  be aligned on the design orbit 
at the entrance of a bunch compression chicane, with a Gaussian longitudinal density
distribution and rms bunch length $\sigma_{s0}$ 
\beq
n(s_{0},t_{0})=n_{0}(\mu) = \frac{1}{\sqrt{2 \pi}\sigma_{s0}}{e^{-\mu^2/2 \sigma_{s0}^2}}.
\eeq
Here we let each electron be identified by the parameter $\mu$, which is its initial longitudinal position 
\beq
s(\mu,t_{0})=s_{0}=\mu  \hspace{0.2in} (s>0 \,\, \mbox{for bunch head}).
\label{mudf}
\eeq
In order to compress the bunch using the chicane, a linear energy correlation was
imposed on the bunch by an upstream RF cavity, along with a slight second-order energy correlation due to the
curvature of the RF wave form. The relative energy deviation from the design energy is then
\beq
\delta(\mu,t_{0})=-\delta_{1}\frac{\mu}{\sigma_{s0}}-\delta_{2}\left(\frac{\mu}{\sigma_{s0}}\right)^2   
(\delta_{1},\delta_{2} >0, \delta_{2} \ll \delta_{1}),
\label{de}
\eeq
where we assume no uncorrelated energy spread.
When the beam  propagates to the end of the chicane at $t=t_{f}$, the final longitudinal coordinates of the electrons
are 
\bea
\lefteqn{ \hspace{-0.3in} s(\mu,t_{f})  = s(\mu,t_{0})+R_{56} \delta(\mu,t_{0})+T_{566} [\delta(\mu,t_{0})]^2} \\
 & & = s_(\mu,t_{0})(1-\frac{R_{56} \delta_{1}}{\sigma_{s0}}) - \alpha [s_(\mu,t_{0})]^2
\eea
with $\alpha \equiv (R_{56} \delta_{2}-T_{566}\delta_{1}^2)/\sigma_{s0}^2$.
One can obtain the maximum compression of the bunch by
choosing the initial bunch length and the initial energy spread to satisfy
\beq
1-R_{56} \delta_{1}/\sigma_{s0}=0, \hspace{0.2in}
s(\mu,t_{f})=s_{f}=- \alpha [s(\mu,t_{0})]^2.
\label{cmpr}
\eeq
For typical bunch compression chicane, one has $R_{56}>0$ and $T_{566}<0$.
Therefore $\alpha>0$, which implies $s_{f} \le 0$ from Eq.~(\ref{cmpr}). 
Using Eqs.~(\ref{cmpr}) and (\ref{mudf}), we have
\beq
\mu(s_{f}) =\sqrt{ -s_{f}/\alpha}  \hspace{0.4in} (\alpha >0, s_{f} \le 0).
\eeq
The final longitudinal density distribution can be obtained from charge conservation
$n_{0}(\mu)d\mu=n(s_{f},t_{f})ds_{f}$:
\bea
& & \hspace{-0.3in}n(s_{f},t_{f}) = \frac{1}{\sqrt{2\pi}\sigma_{sf}} 
\frac{e^{s_{f}/\sqrt{2}\sigma_{sf}}}{\sqrt{-s_{f}/\sqrt{2}\sigma_{sf}}} H(-s_{f}), \label{n0} \\
& & \hspace{-0.3in} \sigma_{sf}=\sqrt{\langle s_{f}^2\rangle-\langle s_{f}\rangle^2}= \sqrt{2}\alpha \sigma_{s0}^{2}.
\label{fe}
\eea
where $H(-s_{f})$ is the Heaviside step function, and $\sigma_{sf}$ is the rms of the
final longitudinal distribution. The final longitudinal phase space distribution can be obtained as
\beq
s_{f} \simeq - (\sigma_{sf}/\sqrt{2}\delta_{1}^2) \delta^{2} 
\label{se} 
\eeq

For example, when $\sigma_{s0}=1.26$ mm, $R_{56}=45$ mm, and $\delta_{1}=0.028$, 
the compression condition Eq.~(\ref{cmpr}) is satisfied. With $\alpha=0.08$ mm$^{-1}$,
Eq.~(\ref{fe}) gives the final compressed bunch length $\sigma_{sf}=0.18$ mm. 
For a  realistic beam, uncorrelated energy spread $\delta_{\mbox{\scriptsize un}}$ should be added to 
Eq.~(\ref{de}) (here we assume  $\delta_{\mbox{\scriptsize un}}$ has a Gaussian distribution with
$\langle \delta_{\mbox{\scriptsize un}} \rangle=0$, and rms width $\delta_{\mbox{\scriptsize un}}^{\mbox
{\scriptsize rms}}$).
As a result, one finds the final rms bunch length satisfies
\beq
\sigma_{s}^{\mbox{\scriptsize eff}}=\sqrt{\langle s_{f}^2\rangle-\langle s_{f}\rangle^2} = 
\sigma_{sf}\sqrt{1 + a^2},
\label{width}
\eeq
with $\sigma_{sf}$ given by Eq.~(\ref{fe}), and $a=R_{56} \delta_{\mbox{\scriptsize un}}/\sigma_{sf}$.
An example of the longitudinal phase space distribution described by Eq.~(\ref{se}), with an additional
width due to  $\delta_{\mbox{\scriptsize un}} \neq 0$ as given by Eq.~(\ref{width}), is shown in Fig.1.

\begin{figure}[h] 
\psfig{figure=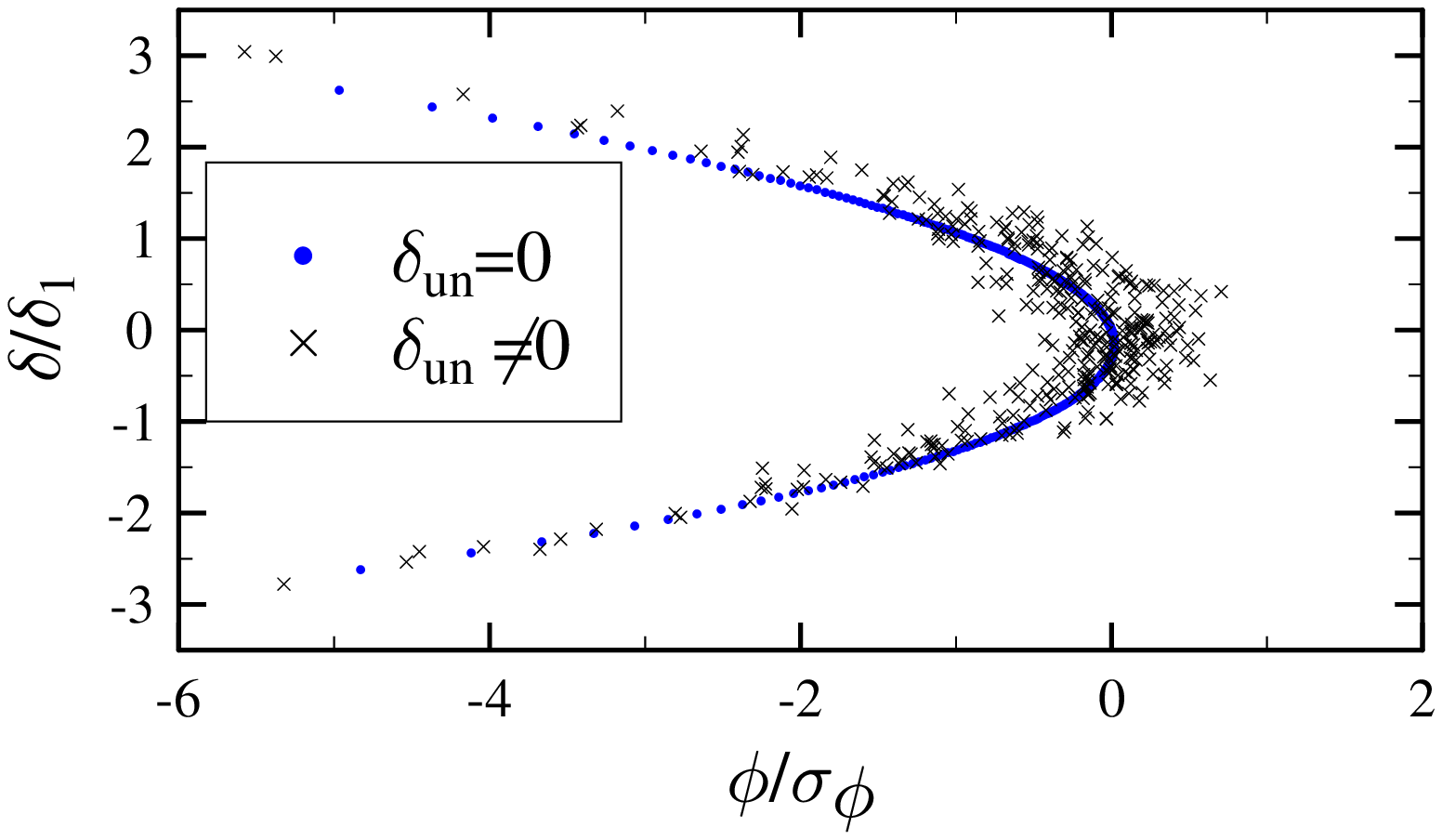,height=1.5in,width=2.8in,bbllx=0.5in,bblly=4.5in,bburx=6.75in ,bbury=7.5in}
\caption{Example of the longitudinal phase space distribution for a compressed beam with RF curvature effect.}
\label{rf}
\end{figure}

\vspace{-0.2in}
\section{CSR for a Compressed Beam}

Next, we study the CSR self-interaction of a rigid-line compressed bunch 
in the steady-state circular motion. The longitudinal density distribution function of the bunch
is $\lambda(\phi)$ for $\phi=s/R$, with the rms angular width $\sigma_{\phi}=\sigma_{s}/R$ for the rms  bunch length $\sigma_{s}$  and the orbit radius $R$.

\subsection{General Formulas}

The longitudinal collective force on the bunch via space-charge and CSR self-interaction is \cite{derbenev, murphy}: 
\bea
& & \hspace{-0.5in}F_{\theta}(\phi)  = \frac{e\partial(\Phi-\mbox{\boldmath $\beta$} \cdot {\bf A})}{\beta c\partial t} \nonumber  \\
& & \hspace{-0.3in}=\frac{-Ne^2}{R^2}\frac{\partial}{\partial \phi}\int_{0}^{\infty} \frac{1-\beta^{2} \cos\theta}{2 \sin(\theta/2)} \lambda(\phi-\varphi)d\varphi
\eea
where $\mbox{\boldmath $\beta$}={\bf v}/c$, $\beta=|\mbox{\boldmath $\beta$}|$, $\gamma=1/\sqrt{1-\beta^2}$, and 
$\theta$ is an implicit function of $\varphi$ via the retardation relation
$\varphi=\theta-2 \beta \sin(\theta/2)$. In this paper, 
we treat only the high-energy case when  $\gamma \gg \theta^{-1}$ and
$\theta \simeq 2(3\varphi)^{1/3}$.
In this case  $F_{\theta}(\phi)$ is dominated by the radiation interaction:
\beq
F_{\theta}(\phi) 
 \simeq \frac{-2Ne^2}{3^{1/3}R^{2}}\int_{0}^{\infty} \varphi^{-1/3}\frac{\partial}{\partial \phi}\lambda(\phi-\varphi)
d\varphi.
\label{fs}
\eeq
The CSR power due to the radiation interaction is
\beq
P = -N \int F_{\theta}(\phi)\lambda(\phi) d\phi.
\label{ps}
\eeq
Results for the longitudinal collective force and the CSR power
for a rigid-line Gaussian bunch are \cite{derbenev, murphy}:
\bea
& & \lambda^{\mbox{\scriptsize Gauss}}(\phi)=\frac{1}{\sqrt{2 \pi} \sigma_{\phi}} e^{-\phi^2/2\sigma_{\phi}^2}
\hspace{0.1in} (\sigma_{\phi} \gg \frac{1}{\gamma^3}), \\
& & \hspace{-0.2in}F_{\theta}^{\mbox{\scriptsize Gauss}}(\phi) \simeq F_{g} g(\phi), \, 
F_{g}=\frac{2 N e^2}{3^{1/3}\sqrt{2 \pi}R^{2}\sigma_{\phi}^{4/3}}, \label{fgs} \\
& & P^{\mbox{\scriptsize Gauss}} \simeq \frac{N^2e^2}{R^{2}\sigma_{\phi}^{4/3}} \, \frac{3^{1/6} \Gamma^2(2/3)}{2 \pi},
\eea
where $\Gamma(x)$ is the Gamma function, and
\beq
g(\phi)=\int_{0}^{\infty} \frac{(\phi/\sigma_{\phi}-\phi_{1})}{\phi_{1}^{1/3}}e^{-(\phi/\sigma_{\phi}-\phi_{1})^2/2}d\phi_{1}.
\eeq

\subsection{CSR Interaction for a Compressed Bunch}

The angular distribution for a compressed bunch $\lambda^{\mbox{\scriptsize cmpr}}(\phi)$ with intrinsic width due to $\delta_{\mbox{\scriptsize un}}\neq 0$ is the convolution of the angular density distribution  $\lambda_{0}^{\mbox{\scriptsize cmpr}}(\phi)$ with  $\delta_{\mbox{\scriptsize un}} = 0$ and a Gaussian distribution $\lambda_{m}(\phi)$:
\bea
& & \hspace{-0.5in}\lambda^{\mbox{\scriptsize cmpr}}(\phi) =\int_{-\infty}^{\infty}\lambda_{0}^{\mbox{\scriptsize cmpr}}(\phi-\varphi)\lambda_{m}(\varphi)d\varphi \label{feff},\\
& & \hspace{-0.5in}\lambda^{\mbox{\scriptsize cmpr}}_{0}(\phi) = \frac{1}{\sqrt{2\pi}\sigma_{\phi}} \ds \frac{e^{\phi/\sqrt{2}\sigma_{\phi}}}{
\sqrt{-\phi/\sqrt{2}\sigma_{\phi}}} H(-\phi), \label{exp} \\
& &\hspace{-0.5in}\lambda_{m}(\phi)=\frac{1}{\sqrt{2\pi}\sigma_{m\phi}} e^{-\phi^{2}/2\sigma_{m\phi}^2},    \, \,  \sigma_{m\phi}=\frac{R_{56}\delta_{\mbox{\scriptsize un}}^{\mbox{\scriptsize rms}}}{R},
\label{fem}  
\eea
where $\lambda^{\mbox{\scriptsize cmpr}}_{0}(\phi)$ is obtained from Eq.~(\ref{n0}).
We then proceed to analyze the longitudinal CSR self-interaction force for a rigid-line bunch with the density function given in Eq.~(\ref{feff})
under the condition $\sigma_{\phi} > \sigma_{m \phi} \gg \gamma^{-3}$. Combining Eq.~(\ref{feff}) with 
Eq.~(\ref{fs}), and denoting $a$ as the intrinsic width of the bunch relative to the rms bunch length
($0<a<1$):
\beq
a=\frac{\sigma_{w}}{\sigma_{s}}  \hspace{0.2in} (\sigma_{w}=R_{56}\delta_{\mbox{\scriptsize un}}^{\mbox{\scriptsize rms}}), \label{aw}
\eeq
one finds the steady-state CSR longitudinal force for a compressed bunch:
\beq
F_{\theta}^{\mbox{\scriptsize cmpr}}(\phi) = \int_{-\infty}^{\infty} F_{\theta 0}^{\mbox{\scriptsize cmpr}}(\varphi) \lambda_{m}(\phi-\varphi)d\varphi.
\label{fcmpr1}
\eeq
It can be shown that $F_{\theta 0}^{\mbox{\scriptsize cmpr}}(\varphi)$ in Eq.~(\ref{fcmpr1}) is
\bea
& & \hspace{-0.4in}\ F_{\theta0}^{\mbox{\scriptsize cmpr}}(\phi) \simeq 
\frac{-2Ne^2}{3^{1/3}R^{2}}\int_{0}^{\infty} \varphi^{-1/3}\frac{\partial}{\partial \phi}\lambda_{0}^{\mbox{\scriptsize cmpr}}(\phi-\varphi)\,d\varphi \nonumber \\
 & & = -2^{1/4}\, F_{g} \, dG(y)/dy   \hspace{0.4in}  (y=\phi/\sigma_{\phi}), 
\label{fcmp0} 
\eea
with $F_{g}$ given in Eq.~(\ref{fgs}), and
\bea
& & \hspace{-0.2in} G(y)= H(-y)\,e^{-|y|/\sqrt{2}} |y|^{1/6}\, \Gamma\left(\frac{2}{3}\right)\Psi \left(\frac{2}{3},\frac{7}{6};\frac{|y|}{\sqrt{2}}\right) \nonumber \\
& & + H(y)\, y^{1/6}\, \Gamma\left(\frac{1}{2}\right) \Psi\left(\frac{1}{2},\frac{7}{6};\frac{y}{\sqrt{2}}\right),
\eea
where $\Psi(a,\gamma;z)$ is the degenerate hypergeometric function
\beq
\Psi(\alpha,\gamma;z)=\frac{1}{\Gamma(\alpha)} \int_{0}^{\infty}e^{-zt}t^{\alpha-1}(1+t)^{\gamma-\alpha-1}dt. 
\nonumber 
\eeq
As a result, we have 
\bea
& &   \hspace{-0.4in}F_{\theta}^{\mbox{\scriptsize cmpr}}(\phi) 
= \frac{2^{1/4}\, F_{g}}{\sqrt{2\pi}\, a^{5/6}} f\left(\frac{\phi}{a\,\sigma_{\phi}};a \right), 
\label{fcmp} \\
& &  \hspace{-0.4in}f(y;a)=\int_{-\infty}^{\infty} G(a\,x)(y-x)\, e^{-(y-x)^2/2}dx.
\eea
Similarly, the radiation power can also be obtained for the compressed bunch using Eq.~(\ref{ps})
with $\lambda^{\mbox{\scriptsize cmpr}}(\phi)$ in Eq.~(\ref{feff}) and
$F_{\theta}^{\mbox{\scriptsize cmpr}}(\phi)$ in Eq.~(\ref{fcmp}), which gives
\bea
& & \hspace{-0.4in}\frac{P^{\mbox{\scriptsize cmpr}}}{ P^{\mbox{\scriptsize Gauss}}}  \simeq 0.75 \, \frac{I(a)}{a^{5/6}}, \\
& & \hspace{-0.4in} I(a) = -\int_{-\infty}^{\infty} f\left(\frac{\phi}{a\,\sigma_{\phi}};a \right) \,\lambda^{\mbox{\scriptsize cmpr}}(\phi) d\phi.
\eea

Numerical integration shows that $|f(y;a)|_{\mbox{\scriptsize max}}$ --- the maximum of $|f(y;a)|$ for 
fixed $a$ --- is insensitive to $a$ for $0 < a < 1$. As a result, 
for a compressed bunch with fixed  $\sigma_{\phi}$, we found from Eq.~(\ref{fcmp}) the amplitude of the CSR force 
$F_{\theta}^{\mbox{\scriptsize cmpr}}(\phi)$ varies with $a^{-5/6}$. Therefore in contrast to the well-known
scaling law $R^{-2/3}\sigma_{s}^{-4/3}$  for the amplitude of the longitudinal CSR force 
for a Gaussian beam, a bunch described by Eq.~(\ref{feff}) has 
$|F_{\theta}^{\mbox{\scriptsize cmpr}}|_{\mbox{\scriptsize max}} \propto 
R^{-2/3}\sigma_{s}^{-1/2}\sigma_{w}^{-5/6}$ with $\sigma_{w}$ in Eq.~(\ref{aw}) 
denoting the intrinsic width of the bunch.
Likewise, for $a$=0.1, 0.2, and 0.5, we found from numerical integration that $I(a) \simeq$ 0.76, 0.90 and 1.02 
respectively, and correspondingly $P^{\mbox{\scriptsize cmpr}}/P^{\mbox{\scriptsize Gauss}} \simeq$ 3.9, 2.6 and 1.4.
This dependence of the amplitude of the CSR force and power on the intrinsic width of the bunch for a 
fixed rms bunch length manifests the sensitivity of the enhancement of the CSR effect on the local charge 
concentration in a longitudinal charge distribution.

In Figs.~2 and 3, we plot the longitudinal density function for various charge distributions with  the
same rms bunch lengths (except the $\sqrt{1+a^2}$ factor in Eq.~(\ref{width})),  
and the longitudinal CSR collective forces associated with the various distributions. The amplitude of
$F_{\theta}^{\mbox{\scriptsize cmpr}}$ in Fig.~3 agrees with the $a^{-5/6}$ dependence in Eq.~(\ref{fcmp}).
Good agreement of the analytical result in Eq.~(\ref{fcmp}) with the simulation result \cite{li} for the 
CSR force along the example distribution in Fig.~1 is shown in Fig.~4.

This work was inspired by the CSR measurement led by H. H. Braun at CERN, and by discussions with the team
during the measurement. The author is grateful for the 
discussions with J. J. Bisognano, and with P. Piot, C. Bohn, D. Douglas, G. Krafft and B. Yunn for the 
CSR measurement at Jefferson Lab. 
This work  was supported by the U.S. DOE Contract No. DE-AC05-84ER40150.

\begin{figure}[h] 
\psfig{figure=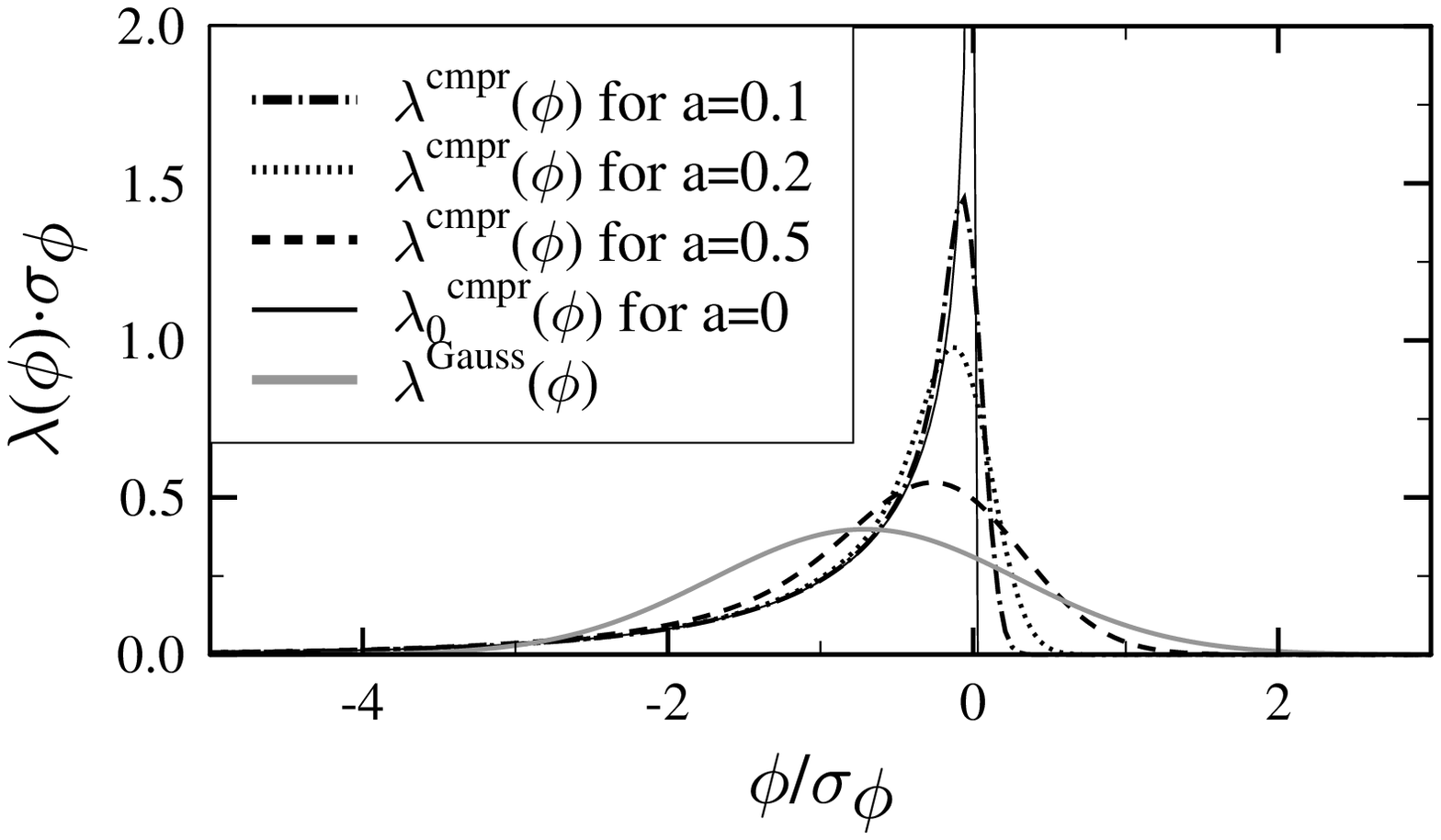,height=1.5in,width=2.8in,bbllx=0.5in,bblly=4.5in,bburx=6.75in ,bbury=7.5in}
\caption{Longitudinal charge distribution for a compressed bunch with intrinsic width described by $a$, compared with 
a Gaussian distribution. All the distributions here have the same angular rms size $\sigma_{\phi}$. }
\label{distribution}
\end{figure}

\begin{figure}[h] 
\psfig{figure=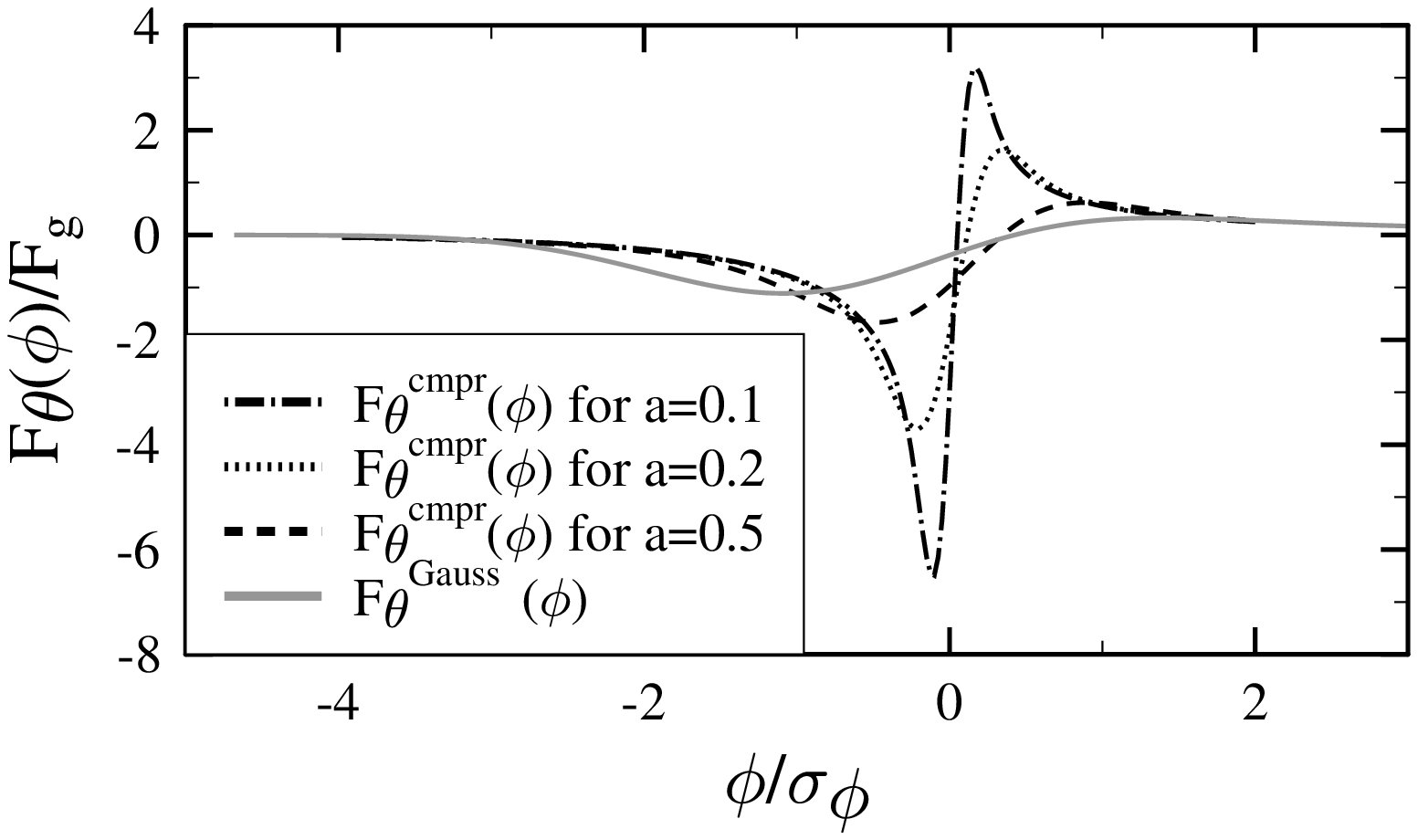,height=1.5in,width=2.8in,bbllx=0.5in,bblly=4.5in,bburx=6.75in ,bbury=7.5in}
\caption{Longitudinal CSR force along the bunch for various charge distributions illustrated in Fig.~2.}
\label{fscsr}
\end{figure}

\begin{figure}[h] 
\psfig{figure=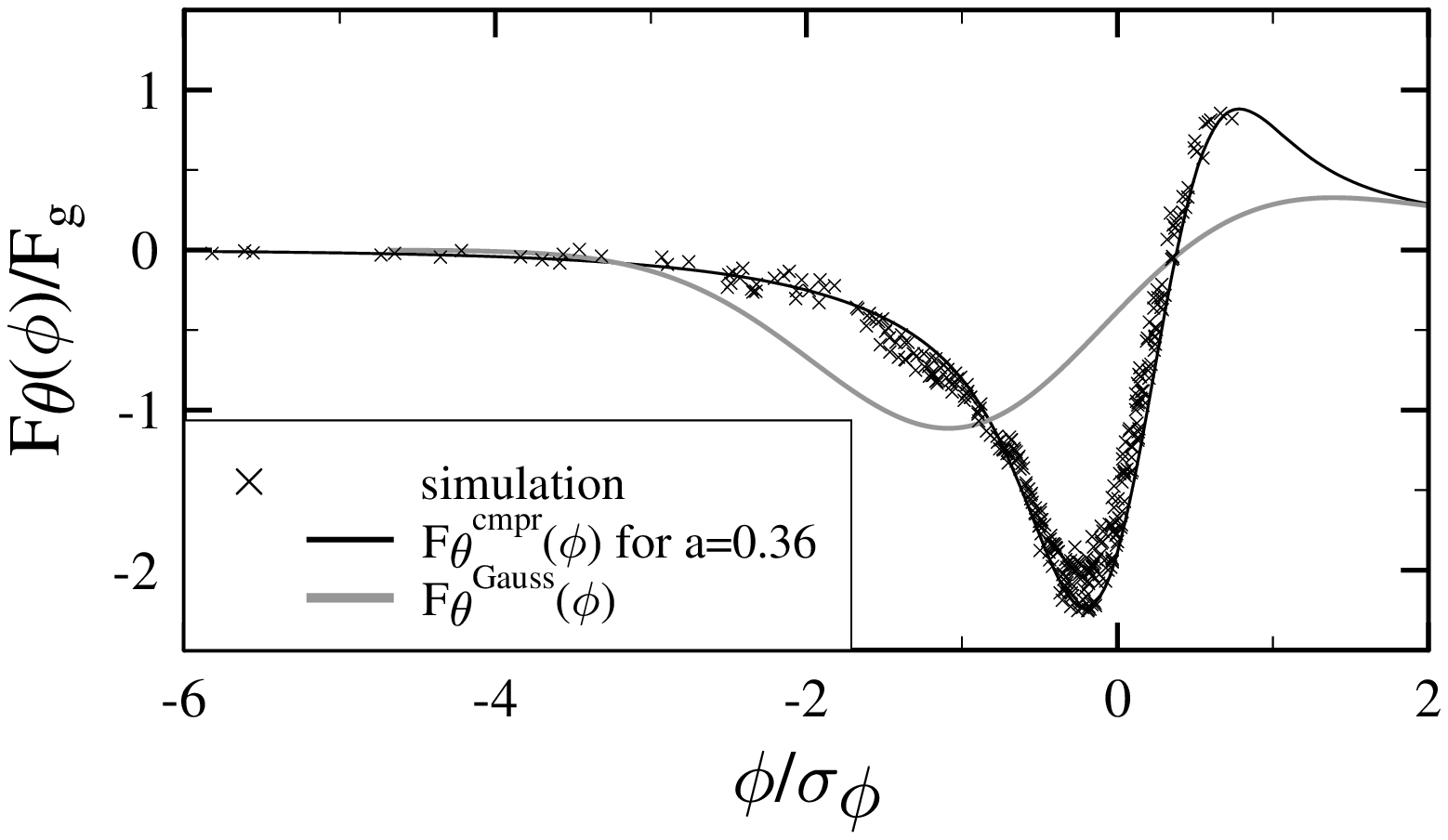,height=1.5in,width=2.8in,bbllx=0.5in,bblly=4.5in,bburx=6.75in ,bbury=7.5in}
\caption{Comparison of the analytical and numerical results of the longitudinal CSR force along the example bunch
illustrated in Fig.~1.  Here we have $\sigma_{x} \simeq 3 \sigma_{s}$.}
\label{fscmpr}
\end{figure}

\end{document}